# Thinging Ethics for Software Engineers

Sabah S. Al-Fedaghi
Computer Engineering Department
Kuwait University
City, Kuwait
sabah.alfedaghi@ku.edu.kw

*Abstract*—Ethical systems are usually described as principles for distinguishing right from wrong and forming beliefs about proper conduct. Ethical topics are complex, with excessively verbose accounts of mental models and intensely ingrained philosophical assumptions. From practical experience, in teaching ethics for software engineering students, an explanation of ethics alone often cannot provide insights of behavior and thought for students. Additionally, it seems that there has been no exploration into the development of a conceptual presentation of ethics that appeals to computer engineers. This is particularly clear in the area of software engineering, which focuses on software and associated tools such as algorithms, diagramming, documentation, modeling and design as applied to various types of data and conceptual artifacts. It seems that software engineers look at ethical materials as a collection of ideas and notions that lack systemization and uniformity. Accordingly, this paper explores a thinging schematization for ethical theories that can serve a role similar to that of modeling languages (e.g., UML). In this approach, thinging means actualization (existence, presence, being) of things and mechanisms that define a boundary around some region of ethically related reality, separating it from everything else. The resultant diagrammatic representation then developed to model the process of making ethical decisions in that region.

*Keywords-ethics; software engineering; conceptual modeling; ethical theory; diagrammatic representation; thinging*

## I. INTRODUCTION

The increasing reliance on computers for infrastructure in modern society has given rise to a host of ethical, social, and legal issues such as those of privacy, intellectual property, and intellectual freedom. These issues have increased and became more complex. Making sound ethical decisions is thus an important subject in computer and software engineering [1]. Computer professional societies (e.g., ACM and IEEE-CS) have proposed several codes of ethics, including the Code of Ethics and Professional Practice for software engineers [2-3] as the standard for teaching and practicing software engineering [4]. The Association for Library and Information Science Education [5] specifies that student learning outcomes in studying ethics include (1) recognizing ethical conflicts; (2) developing responsibility for the consequences of individual and collective actions; and (3) ethical reflection, critical thinking, and the ability to use ethics in professional life.

The field of computer ethics is changing rapidly as concerns over the impact of information and communication technology on society mount. New problems of ethics emerge one after another, and old problems show themselves in new forms. Ethics is often tied to legal procedures and policies that, if breached, can put an organization in the midst of trouble [6].

In this context, we adopt the classical engineering method of *explaining* a phenomenon through modeling. For example, in developing system requirements, a model-based approach is used to depict a system graphically at various levels of granularity and complexity. The resultant unified, conceptual model facilitates communication among different stakeholders such as managers, engineers, and contractors and establishes a uniform vocabulary that leads to common understanding and mental pictures of different states of the system. Similarly, in teaching *modeling* and *explaining* are two closely related practices and "multiple models and representations of concepts" [7] are used to show students how to solve a problem or interpret a text. Representations and models are used in building student understanding [7].

This paper proposes diagrammatically modeling decision making in ethical systems based on the framework of *thinging* wherein things *thing* (a verb that refers to "manifest themselves in the system of concern"), then complete their life cycles though processing, transferring and receiving. The model includes the things and their machines that create, process, release, transfer and receive things.

Ethical systems are usually described as principles for distinguishing right from wrong and beliefs about proper conduct. Ethical topics are complex, with excessively verbose accounts of mental models and intensely ingrained philosophical assumptions. For example, some studies have found that many IT students are unable to distinguish criminal actions from unethical behavior [8].

From practical experience of teaching computer ethics to computer/software engineers (text is Johnson's "Computer Ethics" [9]), it is observed that a textual explanation of ethics often cannot provide insights of ethical behavior and thought for students. Additionally, it seems that there has been no exploration into the style of an ethical *conceptual model* that appeals to computer and software engineers. A conceptual model is an abstraction that describes things of interest to systems. It provides an exploratory basis for understanding and explanation of the phenomenon under consideration. The model can be used as a common representation to focus communication and dialogue, especially in pedagogic environments.

Specifically, we aim to target software engineering students, as software professionals have the power to do good or bad for society and we need to use their knowledge and skills for the benefit of the society [10]. According to the IEEE Computer Society and Association for Computing Machinery (ACM) code of ethics [2-3], every software professional has *obligations* to society, self, profession, product, and employer. Software engineering involves such topics as computer programs and their associated tools such as algorithms, diagramming, documentation, modeling, and designing as applied to various types of data and conceptual artifacts. From other directions and as observed in actual experience of teaching ethics, it seems that software engineers look at ethics as an "alien" topic that contains a collection of ideas and notions that lack systemization and uniformity. Systemization here refers to systems (a highly regarded engineering notion) and their notations, as in representing a system in terms of the classical input-process-output (IPO) model.

Accordingly, this paper explores a thinging-based schematization for ethical theories by expressing ethics in a familiar style for software engineers that is similar to modeling languages (e.g., UML [11]). Since the ability to make diagrams is a valuable and a common skill for programmers skill in software engineering, the proposed method aims at improving students' abilities to describe principles of ethics, to apply a model for ethical decision-making, and to practice diagrammatic communication activities.

Note that this is not a paper in the field of ethics; rather, it introduces a diagrammatic language to describe ethical notions. Consequently, the ethics theories that will be given, if they include errors from the point of ethics, can be corrected by modifying the diagrams without affecting the aim of the paper.

In the next section, we will briefly explain our main tool of modeling, called *Thing Machines* (TM) [12-21]. The example in the section is a new contribution. In Section 3, a brief description of ethics theories is introduced. Sections 4 and 5 give two sample applications of TM. In Section 4, we apply TM to the ethical system of Kantism. We show that TM representation provides a new method of utilizing diagrams to analyze ethics. Motivated by the teaching environment at Kuwait University, we also apply TM to Islamic ethics.

## II. THINGING MACHINES

In philosophy, Thinging refers to "defining a boundary around some portion of reality, separating it from everything else, and then labeling that portion of reality with a name" [22]. According to Heidegger [23], to understand the thingness of things, one needs to reflect on how thinging expresses how a "thing things" that is "gathering", uniting, or tying together its constituents, just as the bridge makes the environment (banks, stream, and landscape) into a unified whole. From slightly different perspectives, thinging and things *thing* (verb) refer to *wujood*: actualization (manifestation), existence, being known or recognized, possession of being, being present, being there, entity (a creature), appearance, opposite of nothingness. For example, a number (in abstract) *has wujood*, but it has no *existence*.

In our approach, there is a strong association between systems and their models. A system is defined through a model. Accordingly, we view an ethical system as a system of "things of ethics." The system also *things* itself by machines of these things. In simple words, as will be exemplified later, it is a web of (abstract) machines represented as a diagram (the grand machine). A machine can thing (create), process, receive, transfer, and/or release other things (see Fig. 1). These "operations" are represented within an abstract Thinging Machine (TM) as shown in Fig. 1.

### A. Example

According to Rosnay [24], the most complete definition of a system is that it is a set of elements in dynamic interaction organized for a goal. Rosnay [24] presents a diagrammatic representation of a reservoir system that fills and empties water that is maintained at the same level as shown in Fig. 2. The figure contains the basic notion that can be used in building the so-called the Thinging Machine (TM) model. Fig. 3 illustrates the notions of things and machines in the reservoir system that will be used in the TM model.

Fig. 4 shows such a system using TM. The water as a thing flows from the outside (circle 1 in the figure) through the valve to the reservoir (2) and outside (3).

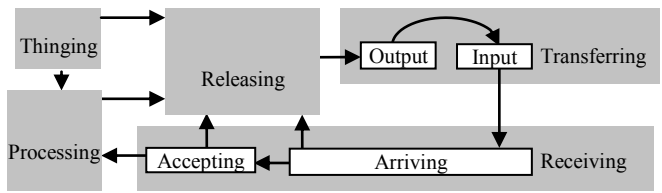

Fig. 1. Thinging machine.

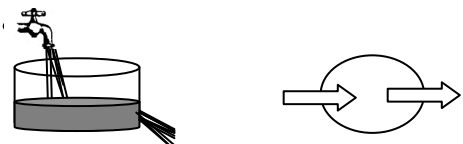

Fig. 2. Diagrammatic representation of a water reservoir system
(Redrawn, partial from Rosnay [24])

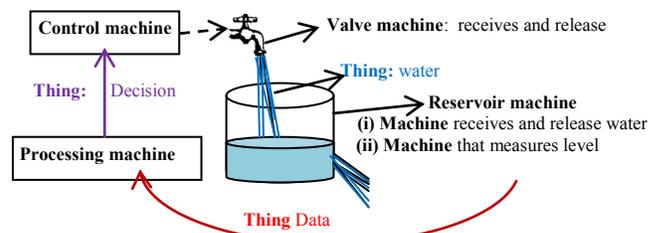

Fig. 3. A water reservoir system and its things/machines.

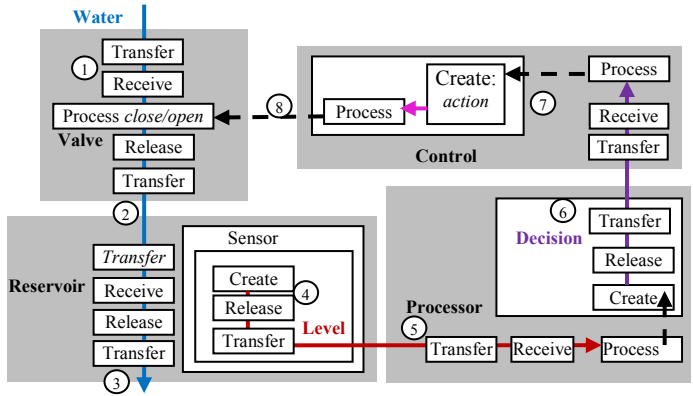

**Fig. 4.** The TM model of the water reservoir system.

The reservoir machine includes a sensor sub-machine that measures the water level (4). The measurement data flow to a processor (5) that triggers a decision machine (the dashed arrow) to generate a control decision (6). The decision flows to control machine of the valve (7), which triggers opening or closing the flow of water from the outside source (8).

An *event* in the TM model is a machine that includes the event itself (sub-machine) and the content of the event, which includes the time sub-machine and region sub-machine of the event at a minimum. Fig. 5 shows the representation of the event *The level measurement is generated and sent to the processor*. Note that the *region* is a sub-graph of the TM static representation of the water reservoir system. For simplicity's sake, we will represent events by only their regions.

Accordingly, Fig. 6 shows meaningful events in the static representation of the reservoir system. Fig. 7 shows the chronology of events that can be used to build a control for the system.

### B. Thinging

We can say that a thing is a *machine* that things (verb), including creating other things that, in turn, are machines that produce things as illustrated in Fig. 8. The chicken is a thing that flows out of the egg. It is also a machine that things (creates), processes (changes), receives, releases, and/or transfers things. In Fig. 8, the chicken machine things (creates) eggs in addition to other things not shown in the figure (e.g., cluck machine, waste machine).

Going by the function of a TM, we define a thing as follows:

*A thing is what manifests itself in the creation, processing, receiving, releasing, and transferring stages of a thinging machine.*

Accordingly, in a TM, we have five kinds of thinging: the machine *creates* (in the sense of *wujood* explained above), *processes (changes)*, *arrives*, *transfers*, and releases (things *wait for departure*). Thinging is the emergence, changing, arriving, departing, and transferring of things.

The utilization of thinging in this paper is not about the philosophical issues related to the ontology of things and their nature; rather, it concerns the representation of things in and machines in a system.

The TM's definition of "thing" broadens its characterization (in comparison to the ontological base of Heidegger [23]'s thinging) by including other thinging aspects: process-ness, receive-ness, transfer-ness, and release-ness. All four features form possible "thingy qualities" [22] after *wujood* (the appearance) of the thing in the grand machine (system of concern).

A thing that has been created refers to a thing that has been born, is acknowledged, exists, appears, or emerges as a separate item in reality or system and with respect to other things. A factory can be a thing that is constructed and inspected as well as a machine that receives other things (e.g., materials) to create products. A factory is a thing when it is processed (e.g., created), and it is a machine when it is processing things (e.g., creating products).

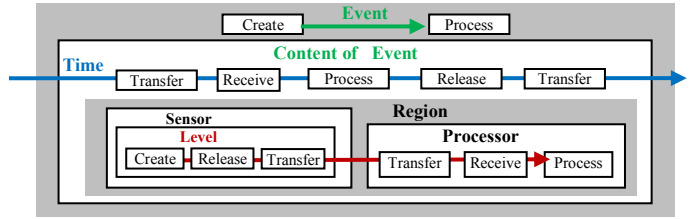

Fig. 5. The event *The level measurement is generated and sent to the processor*.

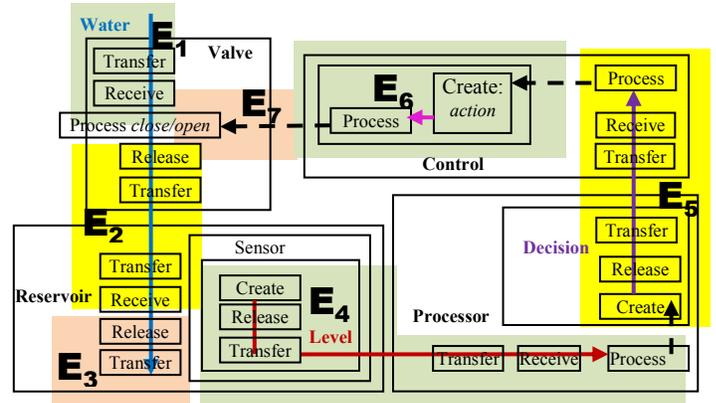

Fig. 6. The TM event-ized representation of the water reservoir.

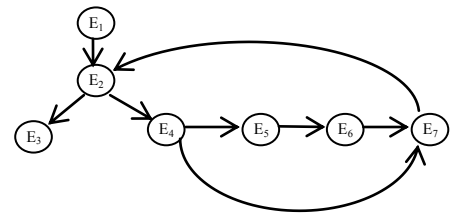

Fig. 7. The chronology of events in the reservoir system.

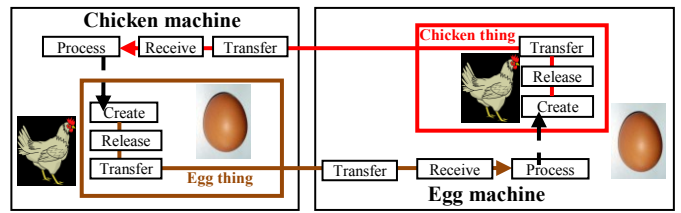

Fig. 8. Illustration of a machine that can be a machine and a machine that can be a thing.

To *Create* a thing means that it *comes about*, and this implies the possibility of its un-thinging (the opposite of *wujood*) within a machine. A collection of machines within a thing forms a larger machine. The stomach is a food-processing machine in the larger digestive machine. The digestive system is one machine in the human being machine whose function is related to the thing 'food.' which is digested (processed) to create waste. A human being is a thing in a school machine.

*Processing* refers to a change that a machine performs on a thing without turning it into a new thing (e.g., a car is processed when its color is changed).

*Receiving* is the flow of a thing to a machine from an outside machine. *Releasing* is exporting a thing outside the machine. It remains within the system, labeled as a released thing, if no release channel is available. *Transferring* is the released thing departing to outside the machine.

The world of a TM consists of an arrangement of machines, wherein each thing has its own unique stream of flow. TM modeling puts together all of the things/machines required to assemble a system (a grand machine).

### III. Ethical Theories

Theories provide explanations to laws. In science, laws manifest regularities in natural phenomena. Scientific laws can be discovered using reason as it is applied to experience. *Natural* laws tell *what is thinging* or *will be thinging*. *Moral* laws are related to *what thinging ought to be*. Instead of natural phenomena, they manifest regularities *in human thinging* (manifestation of oneself or behavior). If human beings were wholly rational beings, moral thinging would be like natural thinging. But since we have inclinations and desires, moral laws appear as imperatives. Nevertheless, there is the claim that morality can be based on natural law. This refers to a system of law that is intrinsic to the structure of the universe [25].

"Being related to ethics" typically refers to thing decisions according to some ethics principles. Normative ethical theories are used to thing ethics judgment when deciding among several alternative courses of response. Ethics determination involves decision-making. Is decision-making in ethics different from other kinds of decision-making, such as in law, rule, policy-making, etc.?

It is often said that moral situations are usually vague regarding which principles are applicable to them. Moral principles may conflict with each other, creating moral dilemmas. Disagreements may also arise about how to interpret and apply these principles in particular situations [26-27]. However, this characteristic of moral situations does not mean that the ethical decision process is fundamentally different from any other decision process that involves vagueness and uncertainly. The decision-making process comprises input, process, and output (IPO mode) and includes factors affecting the determination.

When an ethics problem presents itself, an evaluation machine involves several things, such as objects, persons, circumstances, events, and acts. Ethics values resulting from such a process depend on these factors [28-30].

Actions are morally right when they comply with a moral principle or duty. Thus, in general we have an intended act and an ethical machine that give an act value to an agent who decides which value he/she chooses.

### IV. Modeling Kantism

Ethical theories are divided according to the nature of moral standards used to decide whether a given conduct is right or wrong. Two main categories of normative theories can be identified: the teleological (consequentialist) theories and deontological theories. "Telos" and "deon" in Greek mean "end" and "that which is obligatory", respectively [31]. Deontology is based on the primacy of duty over consequences, where some of which are morally obligatory. Obligation is not necessarily a deontological characteristic. A utilitarian theory, for example, may utilize the concept of obligation teleologically (see [32]). Actions are morally right when they comply with a moral principle or duty. In this paper, we exemplify the application of TM to ethical decision making to two systems, Kantism and Isla.

#### A. Applying TM to Kantism

In Kantism, moral obligations must be carried without qualification, and these must hold for everyone without exception. Hence, the form that moral principles must take is law-like, which can provide the basis for morality. See [28], [33-34]. Here, the will is the human capability to make a decision based on reason. Thus, we should act according to rules that we can embrace as universal laws. Moral principles are categorically (without regard to consequences or exceptions) binding. Humans as rational beings are also moral beings who understand what it takes to live as such. Hence, we impose morality on ourselves, and no one else, as in the case where no one can force us to be rational. As we freely choose to be rational and accept rationality, we also freely choose to be moral and accept morality.

The basic principle categorical imperative (CI) is a modification of the Golden Rule [35] as follows:

*Take an action if its maxim (general principle of conduct) were to become a universal law through your will; i.e., you want others to treat it as a moral law.*

Maxims, according to Kant, are subjective rules that guide action. An ethical decision is universal, applied consistently across time, cultures, and societal norms.

According to Pascal [36], Kantian ethics have great influence on many thinkers, such as Habermas [37], who proposed that action should be based on communication and Rawls' social contract theory [38]. Pascal [36] modelled the "categorical imperative" as shown in Fig. 9. McKnight [39] in her "The Categorical Imperative for Dummies" uses another type of diagram, as shown in Fig. 10.

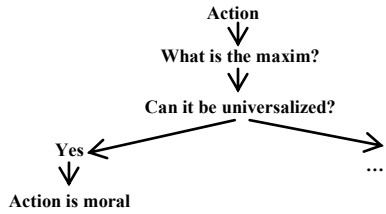

Fig. 9. Categorical Imperative model (Re-drawn, partial from Pascal [36]).

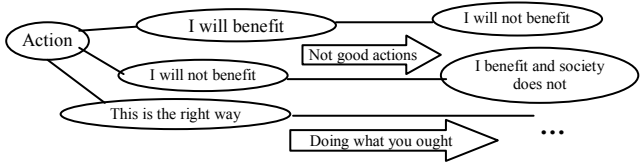

Fig. 10. Kant's theory of duty (re-drawn, partial from McKnight [39]).

According to the Moral Robots blog [40], an action is morally right if it has a good motivation and conforms to Kant's categorical imperative, as explained in Fig. 11. UML use case diagrams [11] have also been used in presenting a method to decide to grant patients' requests for access to their health information based on Health Information Privacy Code, and national and international codes of ethics [41].

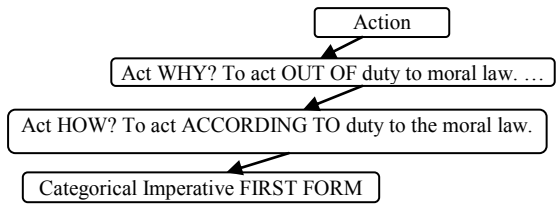

Fig. 11. Kant's Ethics (re-drawn, partial from Moral Robots [40]).

These types of diagrams point to the need for more systematic diagrammatic methods to facilitate explaining ethics. *Systematic* here refers to following a defined methodological approach and an explicitly defined process [42].

Fig. 12 shows the TM representation of Kant's categorical imperative: *Take an action if its maxim were to become a universal law through your will*, and Fig. 13 shows the corresponding dynamic model of such a method of judging actions ethically.

In Fig 12, a person things (i.e., creates) an intended action in his/her mind (circle 1), and this triggers thinging a universe (2) that includes him/herself and others (3—at least two other persons). Each person, I, other-1 and other-2 creates the intended action that flows to the other two in the universe. Processing such a universe (4) triggers the *will* to be in the state of agreeing/disagreeing with such a universe. If the person wants others to treat the intended action as a moral law, then he/she would process the action (5 - determine how to realize the action) and then implement it (6). The copy model (7) guarantees that the mental universe is feasible in reality.

Events in Fig. 13 are created based on meaningfulness to the modeler. Note the time and space machines at the bottom of the figure. The time of the mental universe is received and processed (takes its course) but it never ends (no release and transfer in such a universe). The execution of the action goes on all the time among I, other-1 and other-2.

Additionally, this happens everywhere. It may be interesting to consider that space itself flows as a thing through the repeated events. Suppose that an airplane flows from one place to another; this is conceptually equivalent to space flowing through the airplane. Instead of fixing the space (i.e., Earth) and moving the plane, we fix the plane and move the space (Earth). In reality, if the airplane were fixed, then when the plane is initially over London, we would find it over New York because the Earth turns around itself. The result is that the plane will be over all cities in its circle around the earth.

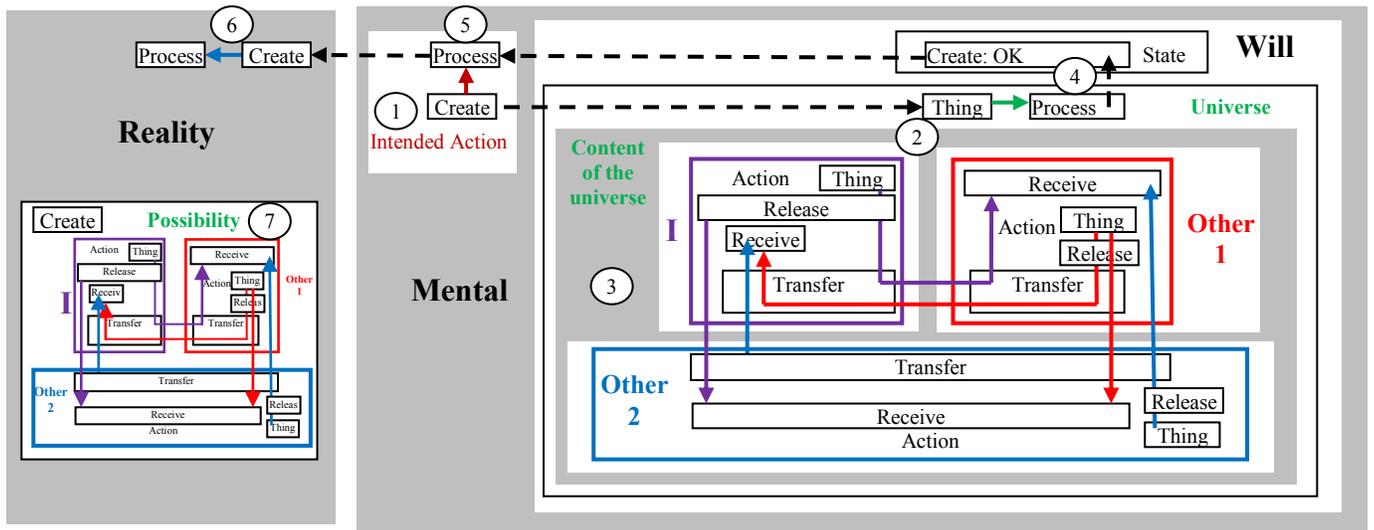

Fig. 12. The TM representation of the categorical imperative.

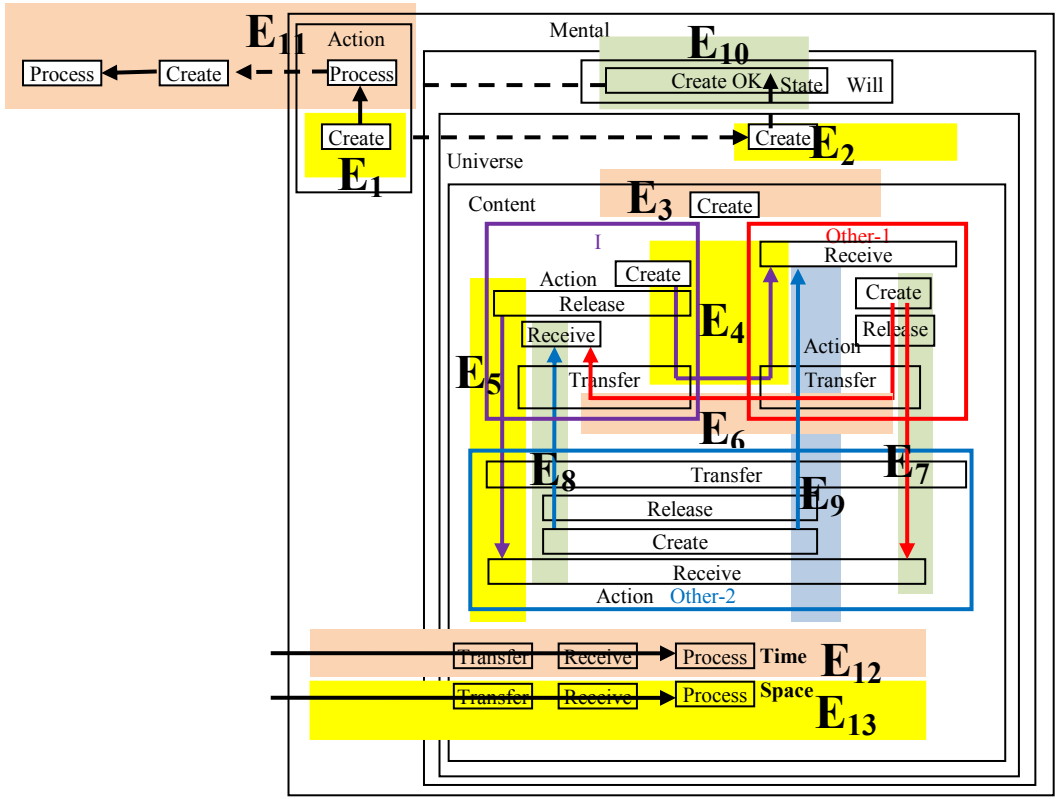

Fig. 13. The TM dynamic representation of the categorical imperative.

Consequently, modeling space (as in the case of modeling time) as flowing in the universe of the figure and never leaving the events means that the events occur everywhere.

Fig. 14 shows the chronology of events, and Fig. 15 illustrates the timing of events wherein some events occur randomly and simultaneously.

*B. Kantism and lying*

One of the major challenges to Kant's reasoning is that it is based on the categorical imperative. Since truth-telling must be universal, one must (if asked) tell a recognized killer the place of his prey. According to Kant, it is one's moral duty to be truthful to a murderer. If he/she is untruthful, then this displays a will to end the practice of thinging the truth. The choice in this case is between assisting a murderer and no *wujood* for truth. Lying is fundamentally wrong, and we cannot thing (create) it even when it eventually triggers good. Untruthfulness means willing universal untruthfulness because the net result is that everyone would thing lies. Also, Kant maintained that if a person performs the correct act, telling the truth, then he/she is not blamable for any outcomes.

Fig. 16 models the situation that *one must (if asked) tell a recognized killer the place of his prey*. It includes person 1 (victim—circle 1), person 2 (the murderer—2) and the person who makes an ethical decision (we will refer to him/her as Agent—3). The victim (person 1) hides in a hiding place (4). This is observed by the agent (5). Additionally, the agent observes the character of person 1 as a victim (6 and 7) and person 2 as a murderer (8 and 9).

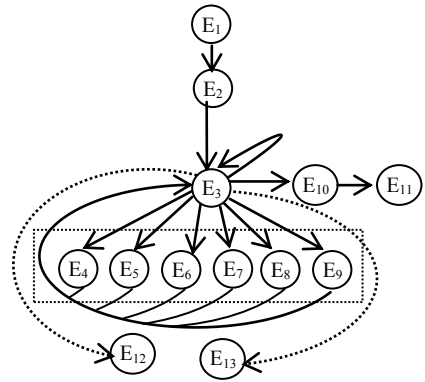

Fig. 14. The chronology of events.

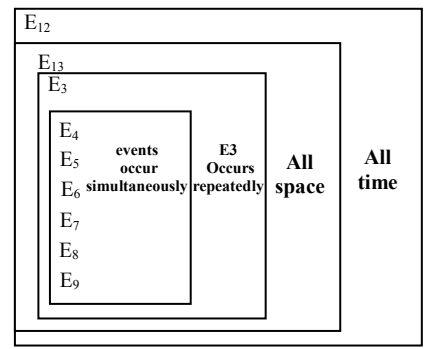

Fig. 15. The timing of events

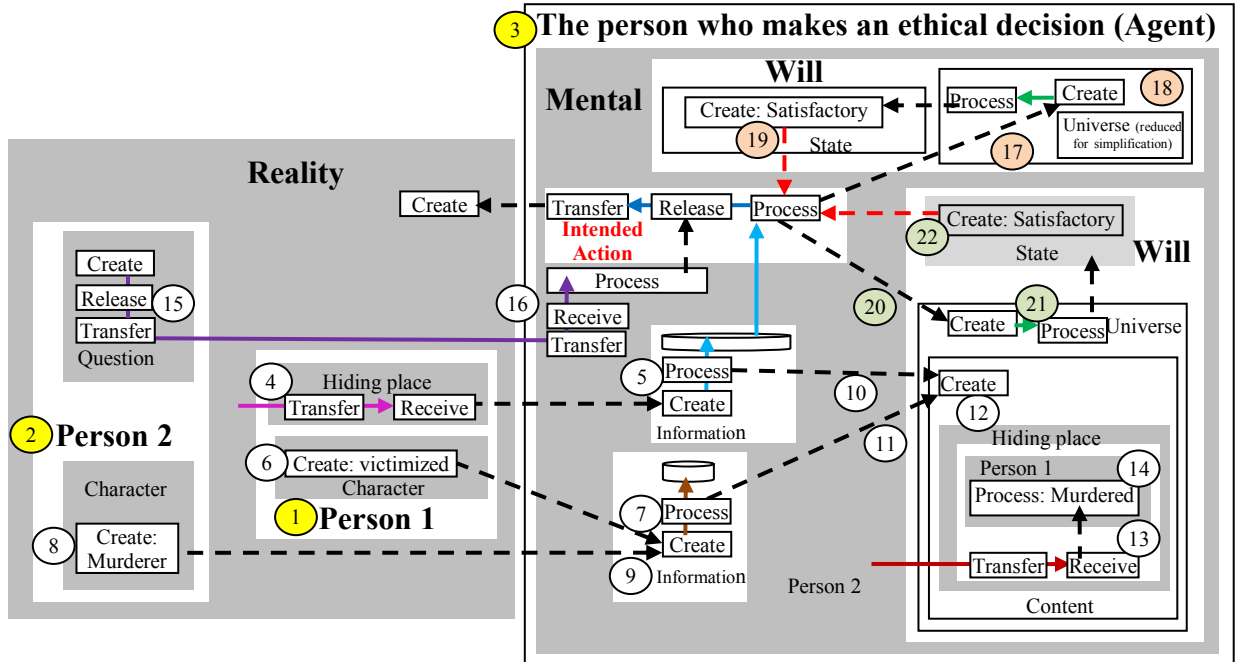

Fig. 16. TM representation of the murderer who pursues a victim dilemma.

From the information about the hiding place (10) and characters of the involved persons (11), the agent creates (12) a mental picture about what will happen if person 2 catches person 1. The picture models the murderer going to the hiding place (13) and murdering the victim (14). Now the murderer asks the agent (15), and the question flows to the agent (16) about the whereabouts of the victim.

Accordingly, the ethical decision to release information about the hideout (15) emerges based on the following factors:
(i) The categorical imperative that triggers the universe (17, 18 and 19), which is "neutral" with respect to the details of the situation. It assumed that the decision is solely based on the categorical imperative. This requires universalizing the act of lying (releasing/transferring misinformation).
(ii) The humanitarian situation as expressed in the mental picture (12). This is triggered by the intended decision (20, 21 and 22) and constructed from the information about the characters of the persons involved (10 and 11). Note that the mental picture appears twice: as imagining what will happen based on information about the characters of the two persons and again as a content of the will. For simplicity's sake, we ignore the appearance of "I" in the universal picture.

Thus, there are two occurrences of universalization: *Everyone is lying* and *every person kills another when given true information about where his/her hideout is*. Thus, the ethical decision to release information about the hideout leads to two contradictory categorical imperatives.

Kant separates individuals from non-individuals when he provides his second formulated principle: the "human integrity" principle. It states: *In every case, treat your own person or that of another, as an end in itself and never merely as a means to an end*. According to Korsgaard [43], "the different formulations [of CI] give different answers to the question of whether if, by lying, someone may prevent a would-be murderer from implementing his/her intentions, that person may do so." From this, it is concluded that different formulations of CI narrow the restrictions imposed by the universalizability requirement.

In our case, we claim that the "human integrity" principle implies that dealing with "information of/about an identified human being" (personal identifiable information) is tantamount to dealing with the human being him/herself.
(iii) There may be other factors, such as the legal issue of assisting a murderer in a crime when telling a recognized killer the place of his prey (not shown in the figure).

The point here is that using TM diagramming facilitates exposing the details of the ethical case. Apparently, the two cases include:
(1) Kant's *pure* categorical imperative as discussed previously, and
(2) Another categorical imperative that involves other details and considerations, as in the case of a murderer pursuing a victim.

Figs. 17 and 18 show different events in this ethical case as follows.

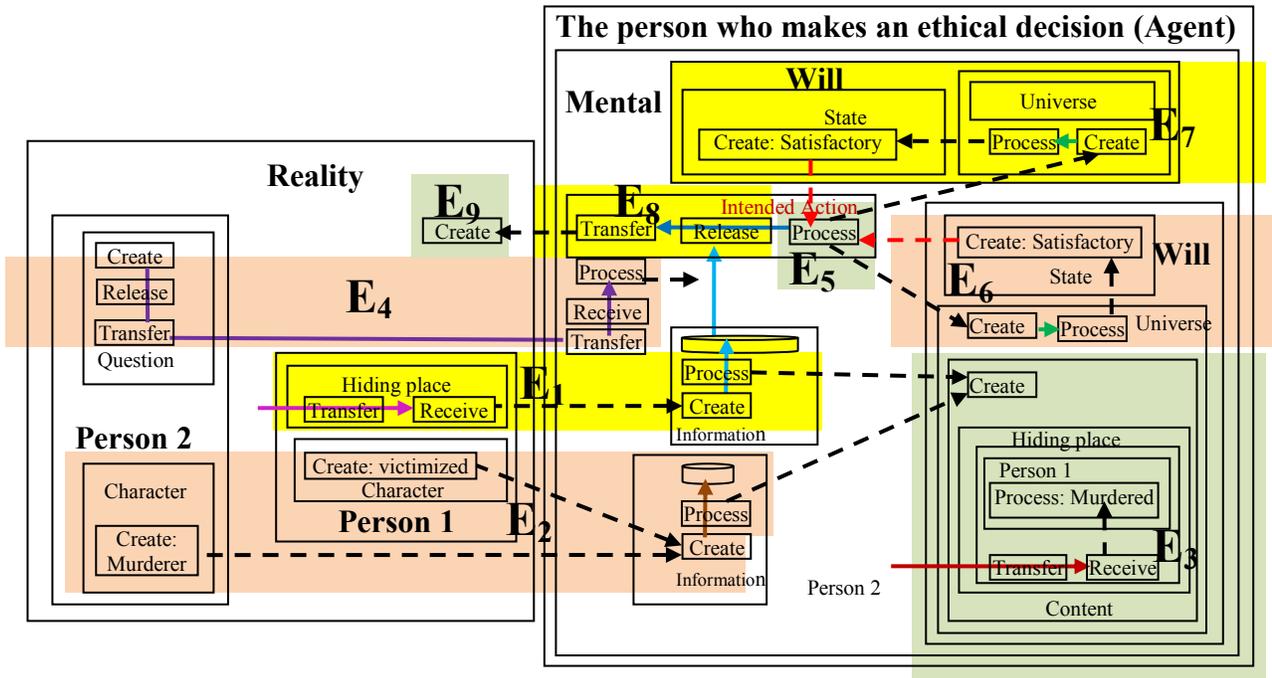

Fig. 17. The TM events of the murderer who pursues a victim.

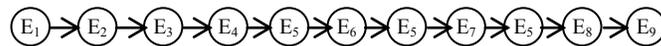

Fig. 18. The chronology of events.

Event 1 ($E_1$): The agent observes person 1 (victim) hiding.
Event 2 ($E_2$): The agent observes the characters of the person to judge that person 1 is a victim and person 2 is a murderer.
Event 3 ($E_3$): From the information in (2), the agent imagines what will happen if the murderer finds the victim.
Event 4 ($E_4$): The murderer askes the agent about the whereabouts of the victim.
Event 5 ($E_5$): The agent processes the intended decision.
Event 6 ($E_6$): He/she thinks: Do I wish to universalize killing?
Event 5 ($E_5$): The agent processes the intended decision again.
Event 7 ($E_7$): He/she thinks: Does he/she will to universalize lying?
Event 8 ($E_8$): The agent makes a decision about releasing information or misinformation.
Event 9 ($E_9$): The agent implements his/her decision.

Such a method of modeling ethical decisions is very suitable for software engineers. It facilities a method of diagramming (e.g., flowcharts, UML [11]) that is familiar in their fields. Note that the aim of this paper is to demonstrate the diagramming tool. Thus, if there is some wrong in the ethical thinking (e.g., imprecise understanding of the categorical imperative), then the diagram can be corrected accordingly.

## V. MODELING ISLAMIC ETHICS

The diagrammatic method described in the previous section has raised the interest of software engineering undergraduate students. However, it is possible to raise their interest further by modeling their own system of ethics. In a multicultural society, different non-secular ethics, such as Islamic, Christian, Judaic, Hindu, and Buddhist ethics can be discussed side-by-side with secular ethics.

This motivates the students because it involves something that is closely related to their personalities. In our school, all students in the computer ethics class are (officially) Muslims. Accordingly, even though there is a lack of knowledge in this area on the part of the instructor, it is possible, with the participation of the students, to develop a reasonably close model of how to make ethical decisions according to Islam. Note that the aim is at using the TM diagrams, so if there is a deviation from the *correct* understanding of Islamic ethics, it easy to modify it.

Additionally, several interpretations of the Islamic ethical system are possible, so selecting the clearest ideas in the literature does not mean adopting these ideas or recommending them for any purpose. This paper discusses an actual experiment in teaching ethics to computer/software engineers with the purpose of exposing them to different approaches and not influencing their ethical thinking.

Accordingly, presenting Islamic ethics was just like presenting Kantism, and selecting Islamic ethics was purely based on the background of the students. For example, in American schools, with multiple students' backgrounds, it is reasonable to model Christian ethics and/or atheist ethics in addition to secular ethics. Note that secularism is not atheism [44].

Theology has a very close relationship with ethics. It includes a history of concern for diverse ethical issues and an important aspect of critical reflection on causes and meanings based on faith and/or a revealed source. Religious ethics are based on divine law. What is right and wrong, what we ought to do or not do, is given by revelation, as moral values and obligations are independent of us. In Islam, moral values and obligations are independent of us; nevertheless, we have complete freedom in selecting to commit ourselves to bringing or not bringing action to the *wujood*.

Actions are judged by intentions, and each action is recompensed according to what a person intends [32].

Fig. 19 shows a model for making an ethical decision in (traditional) Islam according to the understanding of the author. Fig. 20 shows the corresponding event-ized diagram.

In Fig. 19, first a person generates an intended action (circle 1) based on his/her best available information/data and real capability to act (2) and rationality and internal capability to act (3). Then this intended action is processed (4) to check whether it agrees with the Islamic principles as given by the Quran and the Sunna (Sunna is the model pattern of the behavior and mode of life of the Prophet). It is possible that the person might consult an expert or clergy regarding the matter at this stage. Accordingly, the result of this processing is a judgment (fatwa). Either (i) the intended act is permitted in Islam (5) or (ii) it is prohibited by Islam (5).

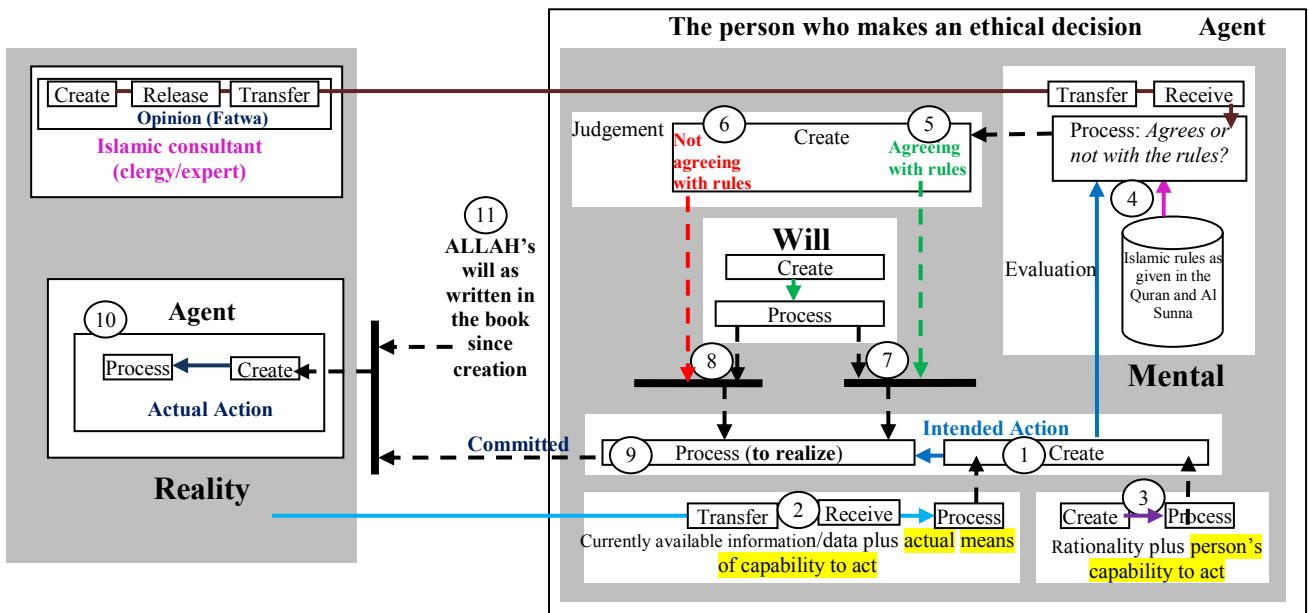

Fig. 19. The TM representation of Islamic ethical decision-making as understood by the author.

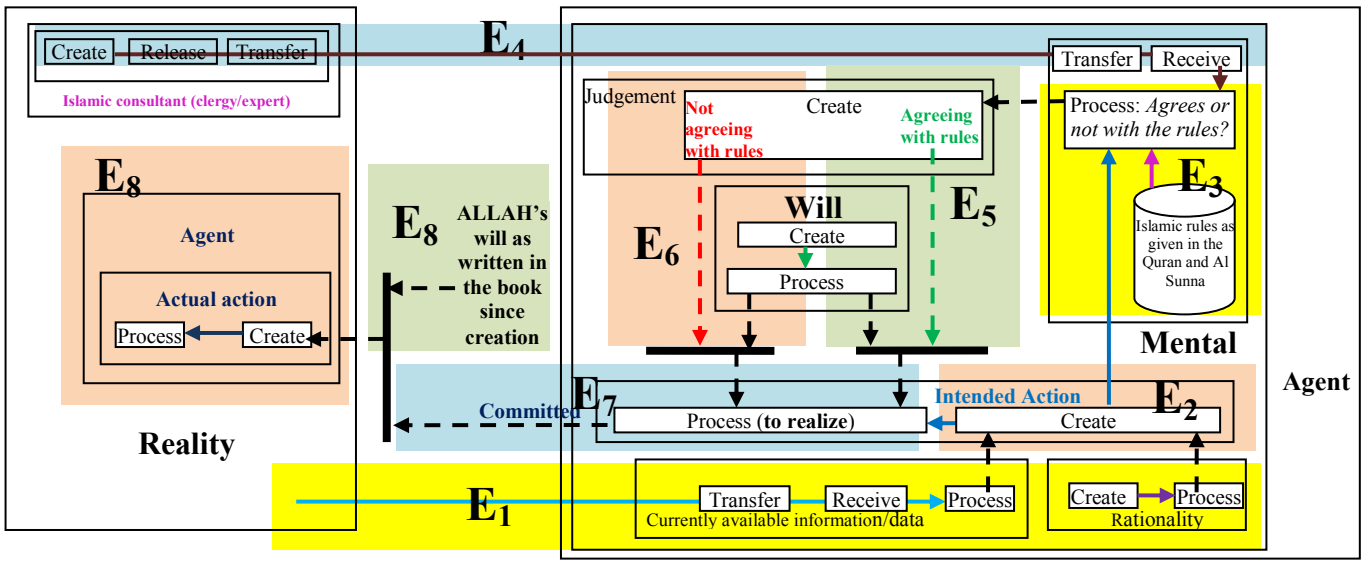

Fig. 20. The TM events of Islamic ethical decision-making as understood by the author.

Then, according to the individual's *free will*, he/she decides to choose which judgement he/she wants to actualize (7 or 8). Consequently, the person "releases" the selected action to actualize it in reality (9). However, in the Islamic faith, all events in reality are written before time (only known by ALLAH) such that no one (even angels) knows what happens until after it has happened. This is called the universal will of ALLAH. Accordingly, an action is actualized in reality (10) only when ALLAH's will (11) coincides with the person's will. This is taken by faith and is an important factor when making an ethical decision as a Muslim.

The set of events of such a scenario (Fig. 20) are as follows.

Event 1 ($E_1$): The agent receives information/data and reasons for the ethical situation or dilemma.

Event 2 ($E_2$): The agent creates an intended action.

Event 3 ($E_3$): The agent applies Islamic rules with regards to his/her intended action.

Event 4 ($E_4$): The agent may consult an expert in Islam.

Event 5 ($E_5$): The agent judges that the action is permitted in Islam.

Event 6 ($E_6$): The agent judges that the action is not permitted in Islam.

Event 7 ($E_7$): The agent takes and actualizes the judgement from (6) or in (7).

Event 8 ($E_8$): ALLAH's will is either to actualize such an action or not.

Event 9 ($E_9$): The action is actualized according to ALLAH's will.

The chronology of these events is shown in Fig. 21.

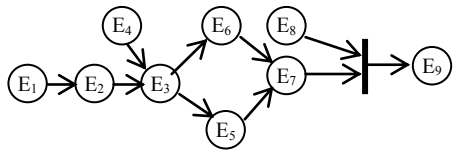

Fig. 21. The chronology of events.

In Kantism, the assumption is that whatever the agent decides by his/her free will is realized in reality; in Islamic ethics, this is not guaranteed because of because of the possible intercession of the will of ALLAH. Note that in Islam, if the an evil action is realized then this does not imply ALLAH's approval (actualizing evil acts) but indicates ALLAH's wisdom in testing humans. If there is no evil in reality then there is no point of free will.

## VI. CONCLUSION

This paper proposes a representational model of ethical systems based on the notions of things and flow. Kantism and Islamic ethics are used as sample ethical systems to demonstrate the applicability of the approach. The modeling technique can be utilized as a pedagogy tool in teaching basic principles of ethics and ethical decision-making. This involves analyzing ethics and design (e.g., robotics). The flowchart-like diagrammatic representation seems to be a familiar style suitable for software engineers.

One weakness of the modeling language is the need to analyze its expressive power of ethical theories and dilemmas that have not been presented in this paper. This is a work for further research.

REFERENCES

[1] G. D. Crnkovic and R. Feldt, "Professional and ethical issues of software engineering curricula: Experiences from a Swedish academic context," in Proceedings of First Workshop on Human Aspects of Software Engineering (HAOSE09), Orlando, Florida, Oct 25, 2009–Oct 26, 2009. Academic Press.

[2] ACM, ACM Code of Ethics and Professional Conduct, June 22nd, 2018. https://www.acm.org/code-of-ethics

[3] IEEE-CS/ACM Joint Task Force on Software Engineering Ethics and Professional Practices, Software Engineering Code of Ethics, Short version, 1999. https://www.google.com/url?sa=t&rct=j&q=&esrc=s&source=web&cd=1&ved=2ahUKEwj46I-Gt8DdAhXChqYKHf30BBEQFjAAegQIABAB&url=https%3A%2F%2Fwww.computer.org%2Fcms%2FComputer.org%2Fprofessional-education%2Fpdf%2Fsoftware-engineering-code-of-ethics.pdf&usg=AOvVaw1_nZEF0dYgHNuiI7iBrJHu

[4] ACM/IEEE Task. Force, "Software engineering code of ethics and professional practice," (version 5.2). https://www.acm.org/about-acm/acm-code-of-ethics-and-professional-conduct

[5] Association for Library and Information Science Education (ALISE). (2007). Position statement on information ethics in LIS education. http://www.alise.org/index.php?option=com_content&view=article&id=51

[6] Status NET, Ethical decision making models and 6 steps of ethical decision making process (No date). https://status.net/articles/ethical-decision-making-process-model-framework/

[7] L. Stickler and G. Sykes, Modeling and explaining content: Definition, research support, and measurement of the ETS® national observational teaching examination (NOTE) assessment series (Research Memorandum No. RM-16-07). Princeton, NJ: Educational Testing Service, 2016.

[8] B. Woodward, Thomas Imboden, "Expansion and validation of the PAPA framework," Information Systems Education Journal (ISEDJ), vol. 9, No. 3, August 2011.

[9] D. G. Johnson, Computer Ethics, 4th edition, Pearson, 2009.

[10] Meherchilakalapudi, Ethical Issues in Software Engineering, blog, March 21, 2009. https://meherchilakalapudi.wordpress.com/2009/03/21/ethical-issues-in-software-engineering/

[11] Object Management Group UML channel, UML & SysML modelling languages: Expertise and blog articles on UML, SysML, and Enterprise Architect modelling tool. http://www.umlchannel.com/en/sysml. [Accessed April 2014]

[12] S. Al-Fedaghi and R. Al-Azmi, "Control of waste water treatment as a flow machine: A case study," 24th IEEE International Conference on Automation and Computing (ICAC'18), 6–7 September 2018, Newcastle University, Newcastle upon Tyne, UK.

[13] S. Al-Fedaghi and M. Bayoumi, "Computer attacks as machines of things that flow," 2018 International Conference on Security and Management (SAM'18), Las Vegas, USA, July 30– August 2, 2018.

[14] S. Al-Fedaghi and N. Al-Huwais, "Toward modeling information in asset management: Case study using Maximo," 4th International Conference on Information Management (ICIM2018), Oxford, UK, May 25–27, 2018.

[15] S. Al-Fedaghi and N. Warsame, "Provenance as a machine," International Conference on Information Society (i-Society), Dublin, Ireland, July 15–18, 2018.

[16] S. Al-Fedaghi and M. Alsharah, "Modeling IT processes: A case study using Microsoft Orchestrator," 4th IEEE International Conference on Advances in Computing and Communication Engineering, Paris, France, June 22–23, 2018.

[17] S. S. Al-Fedaghi, "Thinging for software engineers," International Journal of Computer Science and Information Security (IJCSIS), Vol. 16, No. 7, July 2018.

[18] S. S. Al-Fedaghi and M. Al-Otaibi, "Conceptual modeling of a procurement process: Case study of RFP for public key infrastructure," International Journal of Advanced Computer Science and Applications (IJACSA), Vol 9, No 1, January 2018.

[19] S. S. Al-Fedaghi, "Privacy things: Systematic approach to privacy and personal identifiable information," International Journal of Computer Science and Information Security (IJCSIS), Vol. 16, No. 2, February 2018.

[20] S. Al-Fedaghi and H. Aljenfawi, "A small company as a thinging machine," 10th International Conference on Information Management and Engineering (ICIME 2018), University of Salford, Manchester, United Kingdom, September 22–24, 2018.

[21] S. Al-Fedaghi and M. Alsharah, "Security processes as machines: A case study," Eighth international conference on Innovative Computing Technology (INTECH 2018), London, United Kingdom, August 15–17, 2018.

[22] J. Carreira, "Philosophy is not a luxury," [blog], March 2, 2011, https://philosophyisnotaluxury.com/2011/03/02/to-thing-a-new-verb/

[23] M. Heidegger, "The thing," in Poetry, Language, Thought, A. Hofstadter, Trans. New York: Harper & Row, 1975, pp. 161–184.

[24] Joël de Rosnay, "The macroscope," Principia Cybernetica Project, Translated by Robert Edwards, Harper & Row, 1979.

[25] K. K. Humphreys, What an Engineer Should Know About Ethics, New York/Basel: Marcel Dekker, Inc., 1999.

[26] J. Ladd, "The quest for a code of professional ethics: An intellectual and moral confusion," in Deborah G.Johnson, Ethical Issues in Engineering, Prentice-Hall, Engelwood Cliffs, 1991, pp. 130-136.

[27] M. W. Martin and R. Schinzinger Ethics in Engineering, Third Edition. New York: McGraw-Hill, 1996.

[28] S. D. Ross, Ideals and Responsibilities: Ethical Judgment and Social Identity. Belmont, CA/Albany NY: Wadsworth Publishing Company, 1998.

[29] P.A. Facione, D. Schere, and T. Attig, Ethics and Society, Englewood Cliffs, NJ: Prentice-Hall, 1991.

[30] A. Sesonske, Value and Obligation: The Foundation of Empiricist Ethical Theory. New York: Oxford University Press, 1964.

[31] B. Rosen, Ethical Theory. Mountain View, California: Mayfield Publishing Company, 1990.

[32] G. F. Hourani, Reason and Tradition in Islamic Ethics. Cambridge/London/ New York: Cambridge University Press, 1985.

[33] P. Edwards, Encyclopedia of Philosophy, Vol. 5, New York: Macmillan, 1967.

[34] C. G. Christians, "Ethical theory in a global setting," in Communication Ethics and Global Change, T. W. Cooper (general editor), New York: Longman Inc., 1989.

[35] M. E. Clark, Ariadne's Thread: The Search for New Modes of Thinking. London: Macmillan Press Ltd., 1989

[36] M. Pascal, My Philosopher: Immanuel Kant, 2015, accessed August 2018. https://mediaethicsafternoon.wordpress.com/2015/02/13/my-philosopher-immanuel-kant/

[37] Jürgen Habermas, Stanford Encyclopedia of Philosophy, 2014. https://plato.stanford.edu/entries/habermas/

[38] John Rawls, Theory of Justice, Cambridge, Mass.: The Belknap Press of Harvard University Press, 1971.

[39] L. McKnight, Immanuel Kant and "the Categorical Imperative" for Dummies, Owlcation, 2016.https://owlcation.com/humanities/Immanuel-Kant-and-The-Categorical-Imperative

[40] Moral Robots, Kant's Ethics, Blog. 2017. https://moral-robots.com/philosophy/briefing-kant/

[41] S. McLachlan, K. Dube, T. Gallagher and J. A. Simmonds, "Supporting preliminary decisions on patient requests for access to health records: An integrated ethical and legal framework," 2015 International Conference on Healthcare Informatics (ICHI), Dallas, TX, USA, Oct. 21–Oct. 23, 2015.

[42] G. Marckmann, H. Schmidt, N. Sofaer and D. Strech, "Putting public health ethics into practice: A systematic framework," Front Public Health. Frontiers in Public Health Vol. 3, No. 23, 2015. doi:10.3389/fpubh.2015.00023.

[43] C. Korsgaard, "The right to lie: Kant on dealing with evil." Philosophy and Public Affairs vol. 15, no. 4, 1986.

[44] J. Berlinerblau, Secularism Is Not Atheism, The Blog, Dec 06, 2017.https://www.huffingtonpost.com/jacques-berlinerblau/secularism-is-not-atheism_b_1699588.html